\documentclass[aps,preprint,showpacs,groupedaddress]{revtex4-1}
\usepackage{amssymb}
\usepackage{mathrsfs}
\usepackage{pstricks}
\usepackage{graphicx}
\usepackage{hyphenat}
\usepackage{hyperref}
\begin{document}

%\preprint{}

\title{\bf All-loop order critical exponents for massless scalar field theory with Lorentz violation in the BPHZ method}


%\author{William C. Vieira}
%\email{william.vieira@gmail.com}
%\affiliation{\it Departamento de F\'\i sica, Universidade Federal do Piau\'\i, 64049-550, Teresina, PI, Brazil}
\author{Paulo R. S. Carvalho}
\email{prscarvalho@ufpi.edu.br}
\affiliation{\it Departamento de F\'\i sica, Universidade Federal do Piau\'\i, 64049-550, Teresina, PI, Brazil}

%\thanks{}


%\date{\today}

\begin{abstract}
We compute analytically the all-loop level critical exponents for a massless thermal Lorentz-violating O($N$) self-interacting $\lambda\phi^{4}$ scalar field theory. For that, we evaluate, firstly explicitly up to next-to-leading order and later in a proof by induction up to any loop level, the respective $\beta$-function and anomalous dimensions in a theory renormalized in the massless BPHZ method, where a reduced set of Feynman diagrams to be calculated is needed. We investigate the effect of the Lorentz violation in the outcome for the critical exponents and present the corresponding mathematical explanation and physical interpretation.
\end{abstract}

%\pacs{11.30.-j; 64.60.ae; 11.30.Cp}
%\keywords{}

\maketitle


\section{Introduction}\label{Introduction}

\par It has been shown that symmetry has been a very efficient and elegant guide in the construction process of physical theories. The Standard Model (SM) of high energy elementary particles is based on the $U(1)\otimes SU(2)\otimes SU(3)$ internal symmetries. Another symmetry, a space-time one, to regard in the making of the SM is the Lorentz symmetry. The former are associated to the internal symmetry space of the fields and the latter to the space where the fields are embedded. In recent years there were many proposals to extend the SM to another realms. One of these attempts culminated in the so called Standard Model Extension (SME), where one of the symmetries, the Lorentz symmetry, is broken \cite{PhysRevD.58.116002,PhysRevD.65.056006,PhysRevD.79.125019,PhysRevD.77.085006}. Symmetry properties are of vital importance also in low energy physics, specifically in phase transitions and critical phenomena. In his pioneering work, Wilson evaluated the loop quantum corrections to the classical Landau critical exponents in a massless Lorentz Invariant (LI) O($N$) self-interacting $\lambda\phi^{4}$ scalar field theory \cite{PhysRevLett.28.240}, by applying a theory invariant under Lorentz symmetry and transformations of the Euclidean $N$-dimensional orthogonal group. The critical exponents can be computed in a field theory approach for this statistical mechanic's problem. This approach is named thermal field theory \cite{LeBellac}. In this approach the Lagrangian density plays the role of the energy of the system in the canonical ensemble and the partition function is similar to the generating functional of $n$-point correlation functions. For magnetic systems, the magnetization is proportional to the mean value of a fluctuating field. As it is known, for rather distinct physical systems as a fluid and a ferromagnet, the critical exponents are the same, showing their universal character. They can be written as functions of their dimension $d$, $N$ and symmetry of some $N$-component order parameter, and if the interactions of their constituents are of short- or long-range type. A few works exploring the dependence of the critical exponents on the parameters with an easier experimental access as $d$ \cite{PhysRevB.86.155112,PhysRevE.71.046112} and $N$ \cite{PhysRevLett.110.141601,Butti2005527,PhysRevB.54.7177} as well as on a less accessible one, i.e. the symmetry of the order parameter \cite{PhysRevE.78.061124,Trugenberger2005509} were investigated. We say that two or more systems belong to the same universality class when their critical behaviors are described by an identical set of critical exponents. In empirical terms, the O($N$) universality class is a generalization of the real models with short-range interactions: Ising ($N=1$), XY ($N=2$), Heisenberg ($N=3$), self-avoiding random walk ($N=0$) and spherical ($N \rightarrow \infty$) \cite{Pelissetto2002549}. As the mass of the field corresponds to the difference between an arbitrary temperature $T$ and the critical one $ T_{c}$, massless and massive theories describe critical and noncritical theories, respectively. The thermal fluctuations responsible by the corrections to the mean field values for the critical exponents are now related to the loop quantum contributions to the anomalous dimensions of the theory. The exponents are then extracted from the scaling properties of the renormalized $n$-point correlation functions near (massive theory) or at (massless theory) the critical point. So, equal results for them must be obtained if they are computed at massless or massive theories renormalized at the same or at different renormalization schemes. In fact, the critical exponents were computed in a few distinct theories renormalized at different renormalization schemes \cite{0295-5075-108-2-21001}. Two of them are associated to the field and composite field anomalous dimensions, namely $\eta$ and $\nu$, respectively. The remaining critical exponents $\gamma$, $\alpha$, $\beta$ and $\delta$ can be evaluated by using four independent scaling relations among them. A critical theory is described by a LI Lagrangian density composed of a LI kinetic operator $\partial^{\mu}\phi\partial_{\mu}\phi$ and a self-interacting $\lambda\phi^{4}$ term. The composite field correlation functions are generated by a term like $t\phi^{2}$, where $\phi^{2}\equiv \phi^{2}(y) = \phi(y)\phi(y)$ is evaluated at the same point of space-time. Terms with higher powers like $(\partial^{\mu}\phi\partial_{\mu}\phi)^{3}$, $\phi^{6}$, $(\partial^{\mu}\phi\partial_{\mu}\phi)^{4}$, $\phi^{8}$... are not included because it is known that they give negligible corrections to the critical behavior of the system and are called irrelevant operators \cite{BF02819916,BF02742601,PhysRevD.7.2927}. Operators with odd powers as $\phi^{3}$, $\phi^{5}$ are ruled out by internal symmetry principles. As the magnetization of the system is invariant under a simultaneous change of all spin orientations, the Lagrangian density must be invariant under $\phi\rightarrow -\phi$. An obvious extension to this theory it to consider its Lorentz-violating (LV) version and to investigate the consequences of this symmetry breaking mechanism on the outcome for the critical exponents. Such a theory is attained if we add to the LI Lagrangian density, a LV kinetic term of the form $K_{\mu\nu}\partial^{\mu}\phi\partial^{\nu}\phi$ \cite{PhysRevD.84.065030}, where the dimensionless constant coefficients $K_{\mu\nu}$ are not invariant under Lorentz transformations, i.e. $K_{\mu\nu}^{\prime} \neq \Lambda_{\mu}^{\rho}\Lambda_{\nu}^{\sigma} K_{\rho\sigma}$. If $|K_{\mu\nu}|\ll 1$, we have a slight violation of the referred symmetry. These coefficients are symmetric ($K_{\mu\nu} = K_{\nu\mu}$) and equal for all $N$ components of the field and preserve the O($N$) symmetry of the internal symmetry space of field. Thus the introduction of the LV kinetic operator makes a connection between the breaking of an internal symmetry of the order parameter and a possible modification of the respective universality class, the O($N$) studied here. The achievement of this work is investigating the consequences of Lorentz symmetry breaking on the evaluation of the critical exponents.     

\par This work is organized as follows: In section \ref{Next-to-leading level critical exponents} we compute analytically, by explicit calculations up to next-to-leading order, the critical exponents for critical thermal Lorentz-violating O($N$) self-interacting $\lambda\phi^{4}$ scalar field theory. In section \ref{All-loop order critical exponents} the results of the section \ref{Next-to-leading level critical exponents} are used as inspiration to prove by induction the corresponding critical exponents up to any loop level. We finalize this work in section \ref{Conclusions} with our conclusions.

\section{Next-to-leading level critical exponents}\label{Next-to-leading level critical exponents}

\par The critical behavior of the system is described by the bare $d$-dimensional Lagrangian density 

\begin{eqnarray}\label{huytrji}
&&\mathcal{L}_{B} = \frac{1}{2}\partial^{\mu}\phi_{B}\partial_{\mu}\phi_{B} + \frac{1}{2}K_{\mu\nu}\partial^{\mu}\phi_{B}\partial^{\nu}\phi_{B} + \frac{\lambda_{B}}{4!}\phi_{B}^{4} + \frac{1}{2}t_{B}\phi_{B}^{2}.
\end{eqnarray} 
The quantities, $\phi_{B}$, $\lambda_{B}$ and $t_{B}$ are the unrenormalized field, coupling constant and composite field source, respectively. In a thermal field theory approach, for getting a renormalized theory it is sufficient to renormalize the primitively bare $\Gamma^{(2)}_{B}$, $\Gamma^{(4)}_{B}$ and $\Gamma^{(2,1)}_{B}$ $n$-point correlation functions or $1$PI vertex parts \cite{Amit,ZinnJustin,BrezinLeGuillouZinnJustin}. When written perturbatively up to next-to-leading order (as the tadpole in a massless theory is null, the two-point function up to next-to-leading order must be expanded up to three-loop level), the unrenormalized $\Gamma^{(2)}_{B}$, $\Gamma^{(4)}_{B}$ and $\Gamma^{(2,1)}_{B}$ vertex parts have the expressions

\begin{eqnarray}\label{gtfrdrdes}
&&\Gamma^{(2)}_{B} = \quad\parbox{12mm}{\includegraphics[scale=1.0]{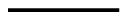}}^{-1} \quad - \quad \frac{1}{6}\hspace{1mm}\parbox{12mm}{\includegraphics[scale=1.0]{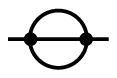}} \quad - \quad \frac{1}{4}\hspace{1mm}\parbox{10mm}{\includegraphics[scale=0.8]{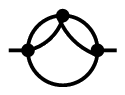}}, 
\end{eqnarray}
\begin{eqnarray}
&&\Gamma^{(4)}_{B} = \quad -\quad \hspace{1mm}\parbox{10mm}{\includegraphics[scale=0.09]{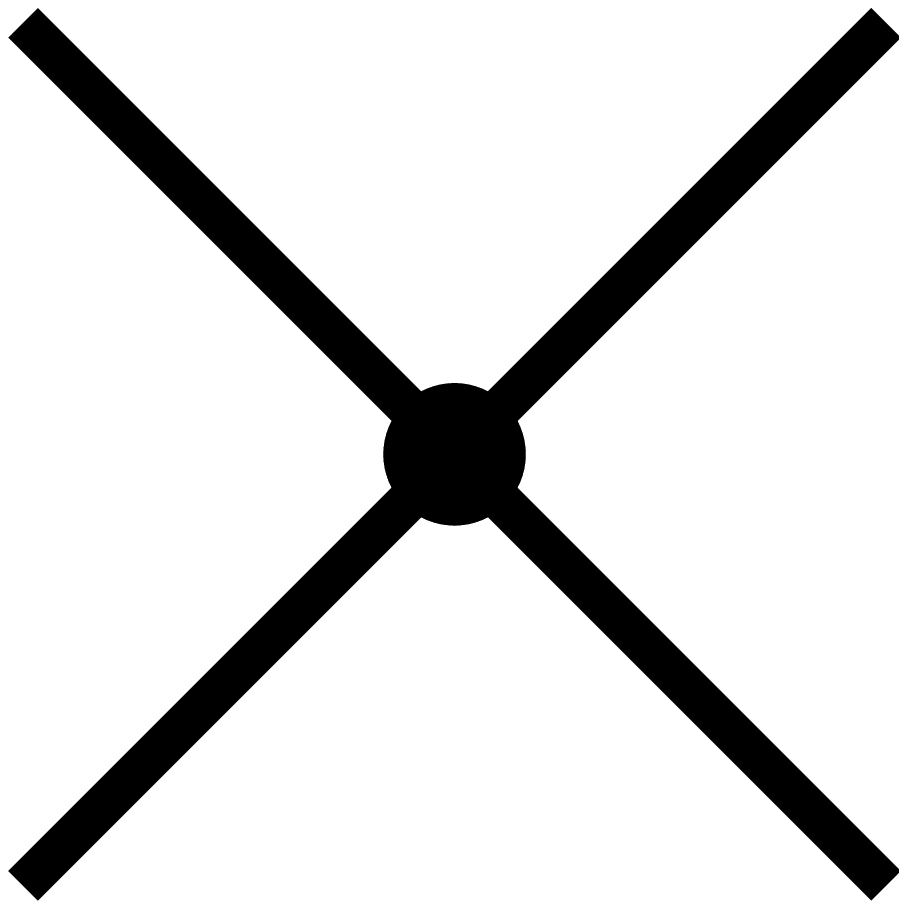}} \quad - \quad \frac{1}{2}\hspace{1mm}\parbox{10mm}{\includegraphics[scale=1.0]{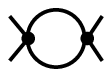}} + 2 \hspace{1mm} perm. \quad - \frac{1}{4}\hspace{1mm}\parbox{16mm}{\includegraphics[scale=1.0]{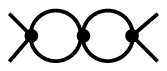}} + 2 \hspace{1mm} perm. \nonumber \\ && - \quad \frac{1}{2}\hspace{1mm}\parbox{12mm}{\includegraphics[scale=0.8]{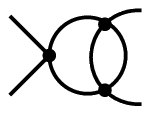}} + 5 \hspace{1mm} perm.,
\end{eqnarray}
\begin{eqnarray}\label{gtfrdesuuji}
&&\Gamma^{(2,1)}_{B} = \quad 1 \quad - \quad \frac{1}{2}\hspace{1mm}\parbox{14mm}{\includegraphics[scale=1.0]{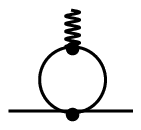}} \quad - \frac{1}{4}\hspace{1mm}\parbox{12mm}{\includegraphics[scale=1.0]{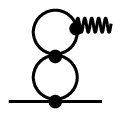}} \quad - \quad \frac{1}{2}\hspace{1mm}\parbox{12mm}{\includegraphics[scale=0.8]{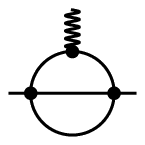}}.
\end{eqnarray}
The theory is renormalized multiplicatively 
\begin{eqnarray}\label{uhygtfrd}
\Gamma^{(n, l)}(P_{i}, Q_{j}, \lambda, \kappa) = Z_{\phi}^{n/2}Z_{\phi^{2}}^{l}\Gamma_{B}^{(n, l)}(P_{i}, Q_{j}, \lambda_{B}, \mu),
\end{eqnarray}
where $\lambda = \mu^{\epsilon}g$ and $g$ are the renormalized dimensional and dimensionless coupling constant, respectively and $\mu$ is an arbitrary momentum scale parameter. In the BPHZ method, the divergences are subtracted by counterterms order by order in perturbation theory. Thus we obtain the renormarlized theory
\begin{eqnarray}\label{huytrjii}
&&\mathcal{L} = \frac{1}{2}Z_{\phi}\partial^{\mu}\phi\partial_{\mu}\phi + \frac{1}{2}K_{\mu\nu}Z_{\phi}\partial^{\mu}\phi\partial^{\nu}\phi + \frac{\mu^{\epsilon}g}{4!}Z_{g}\phi^{4} + \frac{1}{2}tZ_{\phi^{2}}\phi^{2}
\end{eqnarray} 
through the eq. \ref{huytrji} with
\begin{eqnarray}\label{huytr}
\phi = Z_{\phi}^{-1/2}\phi_{B}, 
\end{eqnarray} 
\begin{eqnarray}\label{huytrooiuy}
g = \mu^{-\epsilon}\frac{Z_{\phi}^{2}}{Z_{g}}\lambda_{B}. 
\end{eqnarray} 
and
\begin{eqnarray}\label{huytroo}
t = \frac{Z_{\phi}}{Z_{\phi^{2}}}t_{B} 
\end{eqnarray} 
The renormalization constants in eq. \ref{huytrjii} are given by
\begin{eqnarray}\label{uhguhfgugu}
Z_{\phi} = 1 + \sum_{i=1}^{\infty} c_{\phi}^{i},
\end{eqnarray} 
\begin{eqnarray}\label{uhguhfgugue}
Z_{g} = 1 + \sum_{i=1}^{\infty} c_{g}^{i},
\end{eqnarray} 
\begin{eqnarray}\label{uhguhfguguj}
Z_{\phi^{2}} = 1 + \sum_{i=1}^{\infty} c_{\phi^{2}}^{i}.
\end{eqnarray} 
The $c_{\phi}^{i}$, $c_{g}^{i}$ and $c_{\phi^{2}}^{i}$ coefficients are the $i$-th loop order renormalization constants for the field, renormalized coupling constant and composite field, respectively. The renormalization constants up to next-to-leading order are given by \cite{Kleinert}

\begin{eqnarray}\label{Zphi}
&&Z_{\phi}(g,\epsilon^{-1}) = 1 + \frac{1}{P^2} \Biggl[ \frac{1}{6} \mathcal{K} 
\left(\parbox{10mm}{\includegraphics[scale=1.0]{fig6.eps}}
\right) S_{\parbox{10mm}{\includegraphics[scale=0.5]{fig6.eps}}} + \frac{1}{4} \mathcal{K} 
\left(\parbox{10mm}{\includegraphics[scale=1.0]{fig7.eps}} \right) S_{\parbox{10mm}{\includegraphics[scale=0.5]{fig7.eps}}} \nonumber \\ && + \frac{1}{3} \mathcal{K}
  \left(\parbox{10mm}{\includegraphics[scale=1.0]{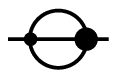}} \right) S_{\parbox{10mm}{\includegraphics[scale=0.5]{fig26.eps}}} \Biggr],
\end{eqnarray}

\begin{eqnarray}\label{Zg}
&&Z_{g}(g,\epsilon^{-1}) = 1 + \frac{1}{\mu^{\epsilon}g} \Biggl[ \frac{1}{2} \mathcal{K} 
\left(\parbox{10mm}{\includegraphics[scale=1.0]{fig10.eps}} + 2 \hspace{1mm} perm.
\right) S_{\parbox{10mm}{\includegraphics[scale=0.5]{fig10.eps}}} + \frac{1}{4} \mathcal{K} 
\left(\parbox{15mm}{\includegraphics[scale=1.0]{fig11.eps}} + 2 \hspace{1mm} perm. \right) S_{\parbox{10mm}{\includegraphics[scale=0.5]{fig11.eps}}} \nonumber \\ && + \frac{1}{2} \mathcal{K} 
\left(\parbox{12mm}{\includegraphics[scale=.8]{fig21.eps}} + 5 \hspace{1mm} perm. \right) S_{\parbox{10mm}{\includegraphics[scale=0.4]{fig21.eps}}} + \mathcal{K}
  \left(\parbox{10mm}{\includegraphics[scale=1.0]{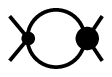}} + 2 \hspace{1mm} perm. \right) S_{\parbox{10mm}{\includegraphics[scale=0.5]{fig25.eps}}}\Biggr],
\end{eqnarray}

\begin{eqnarray}\label{Zphi2}
&&Z_{\phi^{2}}(g,\epsilon^{-1}) = 1 + \frac{1}{2} \mathcal{K} 
\left(\parbox{14mm}{\includegraphics[scale=1.0]{fig14.eps}} \right) S_{\parbox{10mm}{\includegraphics[scale=0.5]{fig14.eps}}} + \frac{1}{4} \mathcal{K} 
\left(\parbox{12mm}{\includegraphics[scale=1.0]{fig16.eps}} \right) S_{\parbox{10mm}{\includegraphics[scale=0.5]{fig16.eps}}} \nonumber \\ && + \frac{1}{2} \mathcal{K} 
\left(\parbox{11mm}{\includegraphics[scale=.8]{fig17.eps}} \right) S_{\parbox{10mm}{\includegraphics[scale=0.4]{fig17.eps}}} \quad + \frac{1}{2} \mathcal{K}
  \left(\parbox{12mm}{\includegraphics[scale=.2]{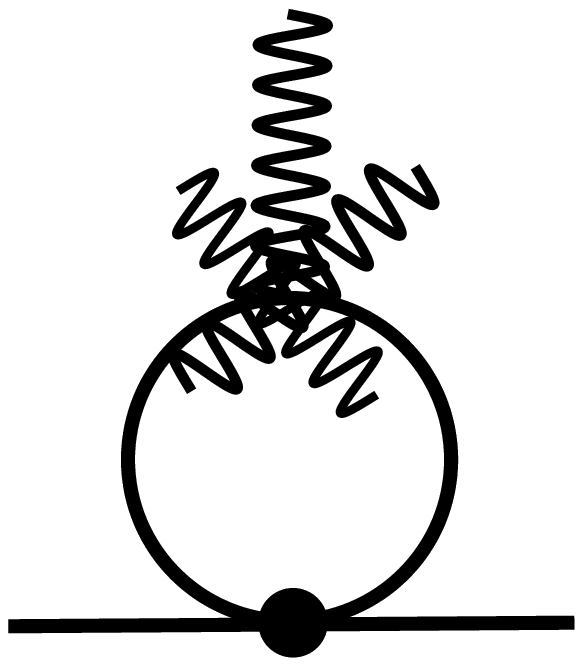}} \right) S_{\parbox{10mm}{\includegraphics[scale=.12]{fig31.eps}}} \quad + \frac{1}{2} \mathcal{K}
  \left(\parbox{12mm}{\includegraphics[scale=.2]{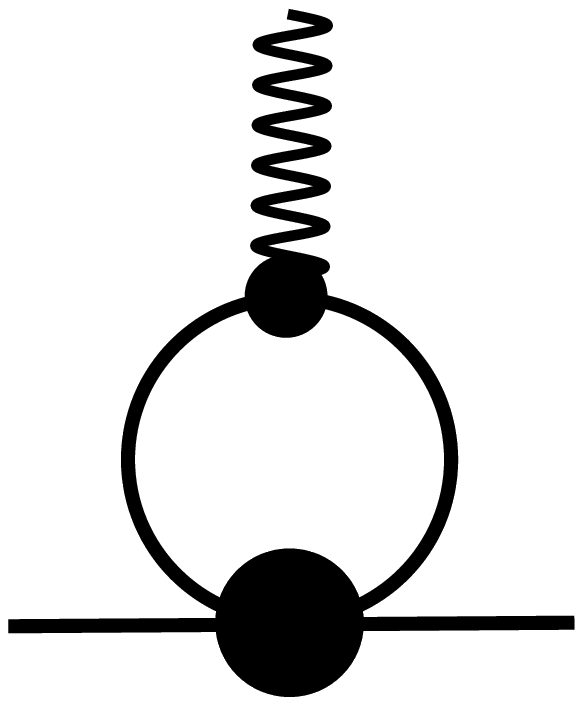}} \right) S_{\parbox{10mm}{\includegraphics[scale=.12]{fig32.eps}}},\nonumber \\
\end{eqnarray}
where $S_{\parbox{6mm}{\includegraphics[scale=0.5]{fig6.eps}}}$ is the symmetry factor for the referred diagram and so on when some $N$-component field is considered. Now we proceed to show how to attain a full solution of the problem by evaluating not all diagrams, but just a few of them. The minimal set of diagrams for that task are
  
\begin{eqnarray}
&&\parbox{12mm}{\includegraphics[scale=1.0]{fig6.eps}} = \lambda^{2}\int \frac{d^{d}q_{1}}{(2\pi)^{d}}\frac{d^{d}q_{2}}{(2\pi)^{d}}\frac{1}{q_{1}^2 + K_{\mu\nu}q_{1}^{\mu}q_{1}^{\nu}}\frac{1}{q_{2}^2 + K_{\mu\nu}q_{2}^{\mu}q_{2}^{\nu}}  \nonumber \\&&  \times\frac{1}{(q_{1} + q_{2} + P)^2 + K_{\mu\nu}(q_{1} + q_{2} + P)^{\mu}(q_{1} + q_{2} + P)^{\nu}}, \nonumber \\
\end{eqnarray}   
\begin{eqnarray}
&&\parbox{12mm}{\includegraphics[scale=0.9]{fig7.eps}} = -\lambda^{3}\int \frac{d^{d}q_{1}}{(2\pi)^{d}}\frac{d^{d}q_{2}}{(2\pi)^{d}}\frac{d^{d}q_{3}}{(2\pi)^{d}}\frac{1}{q_{1}^2 + K_{\mu\nu}q_{1}^{\mu}q_{1}^{\nu}}\frac{1}{q_{2}^2 + K_{\mu\nu}q_{2}^{\mu}q_{2}^{\nu}}\frac{1}{q_{3}^2 + K_{\mu\nu}q_{3}^{\mu}q_{3}^{\nu}} \nonumber \\ &&\times\frac{1}{(q_{1} + q_{2} + P)^2 + K_{\mu\nu}(q_{1} + q_{2} + P)^{\mu}(q_{1} + q_{2} + P)^{\nu}} \nonumber \\ &&\frac{1}{(q_{1} + q_{3} + P)^2 + K_{\mu\nu}(q_{1} + q_{3} + P)^{\mu}(q_{1} + q_{3} + P)^{\nu}}, \nonumber \\
\end{eqnarray} 
\begin{eqnarray}
&&\parbox{12mm}{\includegraphics[scale=1.0]{fig10.eps}} = \lambda^{2}\int \frac{d^{d}q}{(2\pi)^{d}}\frac{1}{q^{2} + K_{\mu\nu}q^{\mu}q^{\nu}}\frac{1}{(q + P)^{2} + K_{\mu\nu}(q + P)^{\mu}(q + P)^{\nu}},
\end{eqnarray} 
where $P_{1} + P_{2} = P$,
\begin{eqnarray}
&&\parbox{14mm}{\includegraphics[scale=0.8]{fig21.eps}} = -\lambda^{3}\int \frac{d^{d}q_{1}}{(2\pi)^{d}}\frac{d^{d}q_{2}}{(2\pi)^{d}}\frac{1}{q_{1}^2 + K_{\mu\nu}q_{1}^{\mu}q_{1}^{\nu}}\frac{1}{(P - q_{1})^{2} + K_{\mu\nu}(P - q_{1})^{\mu}(P - q_{1})^{\nu}} \nonumber \\ && \times\frac{1}{q_{2}^2 + K_{\mu\nu}q_{2}^{\mu}q_{2}^{\nu}}\frac{1}{(q_{1} - q_{2} + P_{3})^2 + K_{\mu\nu}(q_{1} - q_{2} + P_{3})^{\mu}(q_{1} - q_{2} + P_{3})^{\nu}}. 
\end{eqnarray}     

\par Writing the free propagator as an expansion in the small coefficients $K_{\mu\nu}$ 
\begin{eqnarray}\label{expansion}
&& \frac{1}{(q^{2} + K_{\mu\nu}q^{\mu}q^{\nu} + m^{2})^{n}} = \frac{1}{(q^{2} + m^{2})^{n}}\left[ 1 - n\frac{K_{\mu\nu}q^{\mu}q^{\nu}}{q^{2} + m^{2}}  +  \frac{n(n+1)}{2!}\frac{K_{\mu\nu}K_{\rho\sigma}q^{\mu}q^{\nu}q^{\rho}q^{\sigma}}{(q^{2} + m^{2})^{2}} + ...\right]\nonumber \\
\end{eqnarray}
and by applying the formulas in the Appendix \ref{Integral formulas in $d$-dimensional Euclidean momentum space} and dimensional regularization and $\epsilon$-expansion tools in $\epsilon = 4 - d$ \cite{Bollini1972566,BF02895558,'tHooft1972189} we obtain
\begin{eqnarray}
&&\parbox{12mm}{\includegraphics[scale=1.0]{fig6.eps}} = -\frac{g^{2}P^{2}}{8\epsilon}\left[ 1 + \frac{1}{4}\epsilon -2\epsilon L_{3}(P) \right]\Pi^{2} + \frac{g^{2}P^{2}}{4}K_{\mu\nu}L_{3}^{\mu\nu}(P),
\end{eqnarray}  
\begin{eqnarray}
&&\parbox{10mm}{\includegraphics[scale=0.9]{fig7.eps}} = \frac{g^{3}P^{2}}{6\epsilon^{2}}\left[ 1 + \frac{1}{2}\epsilon -3\epsilon L_{3}(P) \right]\Pi^{3} - \frac{g^{3}P^{2}}{2\epsilon}K_{\mu\nu}L_{3}^{\mu\nu}(P),
\end{eqnarray}   
\begin{eqnarray}
&&\parbox{10mm}{\includegraphics[scale=1.0]{fig10.eps}} = \frac{\mu^{\epsilon}g^{2}}{\epsilon}\left\{ \left[1 - \frac{1}{2}\epsilon - \frac{1}{2}\epsilon L(P) \right]\Pi -\frac{1}{2}\epsilon K_{\mu\nu}L^{\mu\nu}(P) \right.  \nonumber \\  &&\left. + \frac{1}{4}\epsilon K_{\mu\nu}K_{\rho\sigma}[L^{\mu\nu}(P)\delta^{\rho\sigma} + L^{\mu\nu\rho\sigma}(P)] \right\},
\end{eqnarray}   
\begin{eqnarray}
&&\parbox{12mm}{\includegraphics[scale=0.8]{fig21.eps}} = -\frac{\mu^{\epsilon}g^{3}}{2\epsilon^{2}}\left\{ \left[1 - \frac{1}{2}\epsilon - \epsilon L(P) \right]\Pi^{2} - \epsilon K_{\mu\nu}L^{\mu\nu}(P) \right.  \nonumber \\  &&\left. + \frac{1}{2}\epsilon K_{\mu\nu}K_{\rho\sigma}[L^{\mu\nu}(P)\delta^{\rho\sigma} + L^{\mu\nu\rho\sigma}(P)] \right\},
\end{eqnarray}  
where
\begin{eqnarray}
\Pi = 1 - \frac{1}{2}K_{\mu\nu}\delta^{\mu\nu} + \frac{1}{8}K_{\mu\nu}K_{\rho\sigma}\delta^{\{\mu\nu}\delta^{\rho\sigma\}} + ...,
\end{eqnarray}
$\delta^{\{\mu\nu}\delta^{\rho\sigma\}} \equiv \delta^{\mu\nu}\delta^{\rho\sigma} + \delta^{\mu\rho}\delta^{\nu\sigma} + \delta^{\mu\sigma}\delta^{\nu\rho}$ and $\delta^{\mu\nu}$ is the Kronecker delta symbol,
\begin{eqnarray}\label{uahuahuahlol}
&&L(P) = \int_{0}^{1}dx\ln\left[\frac{x(1-x)P^{2}}{\mu^{2}}\right],
\end{eqnarray}
\begin{eqnarray}
&&L^{\mu\nu}(P) = \int_{0}^{1}dx\frac{x(1-x)P^{\mu}P^{\nu}}{x(1-x)P^{2}},
\end{eqnarray}
\begin{eqnarray}
&&L_{3}(P) = \int_{0}^{1}dx(1-x)\ln\left[\frac{x(1-x)P^{2}}{\mu^{2}}\right],
\end{eqnarray}
\begin{eqnarray}
&&L_{3}^{\mu\nu}(P) = \int_{0}^{1}dx\frac{x(1-x)^{2}P^{\mu}P^{\nu}}{x(1-x)P^{2}},
\end{eqnarray}
\begin{eqnarray}\label{guigjogkghl}
&&L^{\mu\nu\rho\sigma}(P) = \int_{0}^{1}dx\frac{x^{2}(1-x)^{2}P^{\mu}P^{\nu}P^{\rho}P^{\sigma}}{[x(1-x)P^{2}]^{2}}.
\end{eqnarray}
The diagrams $\parbox{5mm}{\includegraphics[scale=.5]{fig10.eps}}$ and $\parbox{7mm}{\includegraphics[scale=.5]{fig21.eps}}$ are evaluated perturbatively in the small parameters $K_{\mu\nu}$ up
to $\mathcal{O}(K^{2})$ and the $\parbox{6mm}{\includegraphics[scale=.5]{fig6.eps}}$ and $\parbox{6mm}{\includegraphics[scale=.5]{fig7.eps}}$ ones up to $\mathcal{O}(K)$. The computation of the later couple of diagrams up to $\mathcal{O}(K^{2})$ would be very tedious \cite{Carvalho2014320,Carvalho2013850}. The others diagrams can be written in terms of the above as

\begin{eqnarray}
&&\parbox{12mm}{\includegraphics[scale=1.0]{fig26.eps}}\equiv \parbox{12mm}{\includegraphics[scale=1.0]{fig6.eps}}\Bigg|_{-\mu^{\epsilon}g\rightarrow -\mu^{\epsilon}gc_{g}^{1}} =  \nonumber \\ && -\frac{3g^{3}P^{2}}{16\epsilon^{2}}\left[ 1 + \frac{1}{4}\epsilon -2\epsilon L_{3}(P) \right]\Pi^{3} + \frac{3g^{3}P^{2}}{8\epsilon}K_{\mu\nu}L_{3}^{\mu\nu}(P),
\end{eqnarray}
\begin{eqnarray}
&&\parbox{16mm}{\includegraphics[scale=1.0]{fig11.eps}} = -\lambda^{3}\int \frac{d^{d}q_{1}}{(2\pi)^{d}}\frac{1}{q_{1}^{2} + K_{\mu\nu}q_{1}^{\mu}q_{1}^{\nu}}\frac{1}{(q_{1} + P)^{2} + K_{\mu\nu}(q_{1} + P)^{\mu}(q_{1} + P)^{\nu}}\nonumber \\&& \times\int \frac{d^{d}q_{2}}{(2\pi)^{d}}\frac{1}{q_{2}^{2} + K_{\mu\nu}q_{2}^{\mu}q_{2}^{\nu}}\frac{1}{(q_{2} + P)^{2} + K_{\mu\nu}(q_{2} + P)^{\mu}(q_{2} + P)^{\nu}}\equiv -\frac{1}{\lambda}\left(\parbox{12mm}{\includegraphics[scale=1.0]{fig10.eps}}\right)^{2} =  \nonumber \\ && -\frac{\mu^{\epsilon}g^{3}}{\epsilon^{2}}\left\{ \left[1 - \epsilon - \epsilon L(P) \right]\Pi^{2} -\epsilon K_{\mu\nu}L^{\mu\nu}(P)  + \frac{1}{2}\epsilon K_{\mu\nu}K_{\rho\sigma}[2L^{\mu\nu}(P)\delta^{\rho\sigma} + L^{\mu\nu\rho\sigma}(P)] \right\},\nonumber \\
\end{eqnarray}
\begin{eqnarray}
&&\parbox{10mm}{\includegraphics[scale=1.0]{fig25.eps}}\equiv \parbox{12mm}{\includegraphics[scale=1.0]{fig10.eps}}\Bigg|_{-\mu^{\epsilon}g\rightarrow -\mu^{\epsilon}gc_{g}^{1}} =  \nonumber \\ &&  \frac{3\mu^{\epsilon}g^{3}}{2\epsilon^{2}}\left\{ \left[1 - \frac{1}{2}\epsilon - \frac{1}{2}\epsilon L(P) \right]\Pi^{2} -\frac{1}{2}\epsilon K_{\mu\nu}L^{\mu\nu}(P) + \frac{1}{4}\epsilon K_{\mu\nu}K_{\rho\sigma}[2L^{\mu\nu}(P)\delta^{\rho\sigma} + L^{\mu\nu\rho\sigma}(P)] \right\},\nonumber \\
\end{eqnarray}
\begin{eqnarray}
&&\parbox{10mm}{\includegraphics[scale=1.0]{fig14.eps}} \quad = \lambda\int \frac{d^{d}q}{(2\pi)^{d}}\frac{1}{q^{2} + K_{\mu\nu}q^{\mu}q^{\nu}}\frac{1}{(q + Q)^{2} + K_{\mu\nu}(q + Q)^{\mu}(q + Q)^{\nu}} \equiv \frac{1}{\lambda} \parbox{12mm}{\includegraphics[scale=1.0]{fig10.eps}}\Bigg|_{P\rightarrow Q} =  \nonumber \\ &&  \frac{g}{\epsilon}\left\{ \left[1 - \frac{1}{2}\epsilon - \frac{1}{2}\epsilon L(Q) \right]\Pi -\frac{1}{2}\epsilon K_{\mu\nu}L^{\mu\nu}(Q) + \frac{1}{4}\epsilon K_{\mu\nu}K_{\rho\sigma}[L^{\mu\nu}(Q)\delta^{\rho\sigma} + L^{\mu\nu\rho\sigma}(Q)] \right\},
\end{eqnarray}
where $Q = Q_{1} + Q_{2}$,
\begin{eqnarray}
&&\parbox{10mm}{\includegraphics[scale=1.0]{fig16.eps}} \quad = -\lambda^{2}\int \frac{d^{d}q_{1}}{(2\pi)^{d}}\frac{1}{q_{1}^{2} + K_{\mu\nu}q_{1}^{\mu}q_{1}^{\nu}}\frac{1}{(q_{1} + Q)^{2} + K_{\mu\nu}(q_{1} + Q)^{\mu}(q_{1} + Q)^{\nu}}\nonumber \\&& \times\int \frac{d^{d}q_{2}}{(2\pi)^{d}}\frac{1}{q_{2}^{2} + K_{\mu\nu}q_{2}^{\mu}q_{2}^{\nu}}\frac{1}{(q_{2} + Q)^{2} + K_{\mu\nu}(q_{2} + Q)^{\mu}(q_{2} + Q)^{\nu}}\nonumber \\&& \equiv \frac{1}{\lambda}\parbox{16mm}{\includegraphics[scale=1.0]{fig11.eps}}\Bigg|_{P\rightarrow Q} =  \nonumber \\ &&  -\frac{g^{2}}{\epsilon^{2}}\left\{ \left[1 - \epsilon - \epsilon L(Q) \right]\Pi^{2} -\epsilon K_{\mu\nu}L^{\mu\nu}(Q) + \frac{1}{2}\epsilon K_{\mu\nu}K_{\rho\sigma}[2L^{\mu\nu}(Q)\delta^{\rho\sigma} + L^{\mu\nu\rho\sigma}(Q)] \right\},\nonumber \\
\end{eqnarray}
\begin{eqnarray}
&&\parbox{10mm}{\includegraphics[scale=.8]{fig17.eps}} \quad = -\lambda^{2} \int \frac{d^{d}q_{1}}{(2\pi)^{d}}\frac{d^{d}q_{2}}{(2\pi)^{d}}\frac{1}{q_{1}^2 + K_{\mu\nu}q_{1}^{\mu}q_{1}^{\nu}}\frac{1}{(Q - q_{1})^{2} + K_{\mu\nu}(Q - q_{1})^{\mu}(Q - q_{1})^{\nu}} \nonumber \\ && \times\frac{1}{q_{2}^2 + K_{\mu\nu}q_{2}^{\mu}q_{2}^{\nu}}\frac{1}{(q_{1} - q_{2} + Q_{3})^2 + K_{\mu\nu}(q_{1} - q_{2} + Q_{3})^{\mu}(q_{1} - q_{2} + Q_{3})^{\nu}} \nonumber \\ && \equiv \frac{1}{\lambda}\parbox{16mm}{\includegraphics[scale=1.0]{fig21.eps}}\Bigg|_{P\rightarrow Q} =  \nonumber \\ &&  -\frac{g^{2}}{2\epsilon^{2}}\left\{ \left[1 - \frac{1}{2}\epsilon - \epsilon L(Q) \right]\Pi^{2} -\epsilon K_{\mu\nu}L^{\mu\nu}(Q) + \frac{1}{2}\epsilon K_{\mu\nu}K_{\rho\sigma}[2L^{\mu\nu}(Q)\delta^{\rho\sigma} + L^{\mu\nu\rho\sigma}(P)] \right\},\nonumber \\
\end{eqnarray}
\begin{eqnarray}
&&\parbox{10mm}{\includegraphics[scale=.2]{fig31.eps}}\quad\equiv \parbox{14mm}{\includegraphics[scale=1.0]{fig14.eps}}\Bigg|_{-g\rightarrow -gc_{\phi^{2}}^{1}} =  \nonumber \\ &&  \frac{g^{2}}{2\epsilon}\left\{ \left[1 - \frac{1}{2}\epsilon - \frac{1}{2}\epsilon L(Q) \right]\Pi^{2} -\frac{1}{2}\epsilon K_{\mu\nu}L^{\mu\nu}(Q) + \frac{1}{4}\epsilon K_{\mu\nu}K_{\rho\sigma}[2L^{\mu\nu}(Q)\delta^{\rho\sigma} + L^{\mu\nu\rho\sigma}(Q)] \right\},\nonumber \\
\end{eqnarray}
\begin{eqnarray}
&&\parbox{10mm}{\includegraphics[scale=.2]{fig32.eps}}\quad\equiv \parbox{14mm}{\includegraphics[scale=1.0]{fig14.eps}}\Bigg|_{-g\rightarrow -gc_{g}^{1}} =  \nonumber \\ &&  \frac{3g^{2}}{2\epsilon^{2}}\left\{ \left[1 - \frac{1}{2}\epsilon - \frac{1}{2}\epsilon L(Q) \right]\Pi^{2} -\frac{1}{2}\epsilon K_{\mu\nu}L^{\mu\nu}(Q) + \frac{1}{4}\epsilon K_{\mu\nu}K_{\rho\sigma}[2L^{\mu\nu}(Q)\delta^{\rho\sigma} + L^{\mu\nu\rho\sigma}(Q)] \right\}.\nonumber \\
\end{eqnarray}
The coefficients $c_{g}^{1}$ and $c_{\phi^{2}}^{1}$ used above are the one-loop corrections to their respective renormalization constants. They arise from the condition that the theory be finite at one-loop level
\begin{eqnarray}
&&\parbox{6mm}{\includegraphics[scale=.1]{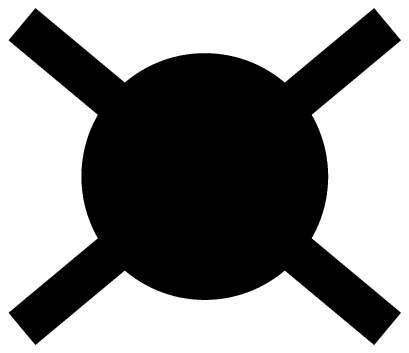}} = -\mu^{\epsilon}gc_{g}^{1} = -\frac{3}{2} \mathcal{K} 
\left(\parbox{10mm}{\includegraphics[scale=1.0]{fig10.eps}} \right),
\end{eqnarray}
\begin{eqnarray}
&&\parbox{16mm}{\includegraphics[scale=.2]{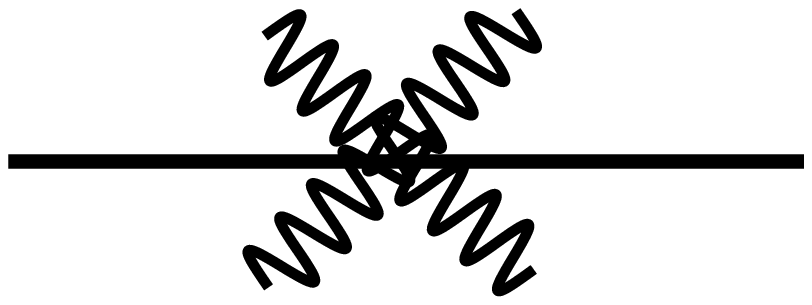}} = -c_{\phi^{2}}^{1} = -\frac{1}{2} \mathcal{K} 
\left(\parbox{11mm}{\includegraphics[scale=.8]{fig14.eps}} \right).
\end{eqnarray}
As the two-point function is finite at one-loop order, $c_{\phi}^{1}=0$. Thus, we obtain for the O($N$) theory the renormalization constants
\begin{eqnarray}\label{uhguhfgugua}
Z_{\phi} = 1 - \frac{N + 2}{144\epsilon}\Pi^{2} g^{2} - \frac{(N + 2)(N + 8)}{1296\epsilon^{2}}\left( 1 - \frac{1}{4}\epsilon \right)\Pi^{3}g^{3},
\end{eqnarray} 
\begin{eqnarray}\label{uhguhfguguea}
Z_{g} = 1 + \frac{N + 8}{6\epsilon}\Pi g + \left[ \frac{(N + 8)^{2}}{36\epsilon^{2}} - \frac{5N + 22}{36\epsilon} \right]\Pi^{2}g^{2},
\end{eqnarray} 
\begin{eqnarray}\label{uhguhfguguja}
Z_{\phi^{2}} = 1 + \frac{N + 2}{6\epsilon}\Pi g + \left[ \frac{(N + 2)(N + 5)}{36\epsilon^{2}} - \frac{N + 2}{24\epsilon} \right]\Pi^{2}g^{2}.
\end{eqnarray} 
Observe that the momentum-dependent integrals in eqs. \ref{uahuahuahlol}-\ref{guigjogkghl} disappeared in the final expressions for the renormalization constants. These constants depend on $K_{\mu\nu}$ only through the LV factor $\Pi$ as a power-law. This fact will be essential for the developments of the next section. The renormalization constants were obtained for arbitrary external momenta, showing the generality of the method. The scaling properties of the $1$PI vertex parts are determined by the renormalization group equations for these correlation functions \cite{Amit}
\begin{eqnarray}\left( \mu\frac{\partial}{\partial\mu} + \beta\frac{\partial}{\partial g} - \frac{1}{2}n\gamma_{\phi} + l\gamma_{\phi^{2}} \right)\Gamma^{(n, l)} = 0
\end{eqnarray}
where 
\begin{eqnarray}\label{kjjffxdzs}
\beta(g) = \mu\frac{\partial g}{\partial \mu},
\end{eqnarray}
\begin{eqnarray}\label{koiuhygtf}
\gamma_{\phi}(g) = \mu\frac{\partial\ln Z_{\phi}}{\partial \mu},
\end{eqnarray}
\begin{eqnarray}
\gamma_{\phi^{2}}(u) = -\mu\frac{\partial\ln Z_{\phi^{2}}}{\partial \mu}.
\end{eqnarray}
Now we evaluate the $\beta$-function and field and composite field anomalous dimensions. They have the following expressions
\begin{eqnarray}\label{reewriretjgjk}
\beta(g) = g\left( -\epsilon + \frac{N + 8}{6}\Pi g - \frac{3N + 14}{12}\Pi^{2}g^{2}\right),
\end{eqnarray} 
\begin{eqnarray}\label{jkjkpfgjrftj}
\gamma_{\phi}(g) = \frac{N + 2}{72}\Pi^{2}g^{2} - \frac{(N + 2)(N + 8)}{1728}\Pi^{3}g^{3},
\end{eqnarray} 
\begin{eqnarray}\label{gfydsguyfsdgufa}
\gamma_{\phi^{2}}(g) = \frac{N + 2}{6}\Pi g\left( 1 - \frac{5}{12}\Pi g \right).
\end{eqnarray} 
The eqs. \ref{reewriretjgjk}-\ref{gfydsguyfsdgufa} are finite functions of $\epsilon$, as promised by the renormalization program. The fixed points of the theory are computed from the eigenvalue condition $\beta(g^{*}) = 0$. The trivial one or the one known as Gaussian is $g^{*} = 0$ and is responsible for the classical Landau values for the critical exponents. For these exponents, the thermal fluctuations at all length scales are not taken into account. These fluctuations are considered when $g$ flows to the nontrivial fixed point. Its value computed from the parentheses in eq. \ref{reewriretjgjk} up to the order of interest is given by
\begin{eqnarray}\label{yagyaguhd}
g^{*} = \frac{6\epsilon}{(N + 8)\Pi}\left\{ 1 + \epsilon\left[ \frac{3(3N + 14)}{(N + 8)^{2}} \right]\right\}\equiv \frac{g^{*(0)}}{\Pi},
\end{eqnarray}
where $g^{*(0)}$ is the LI nontrivial fixed point \cite{Amit}. Thus by applying the relations between the critical exponents $\eta$, $\nu$ and the field and composite field anomalous dimensions $\eta\equiv\gamma_{\phi}(g^{*})$, $\nu^{-1}\equiv 2 - \gamma_{\phi^{2}}(g^{*})$, respectively, the LV $\Pi$ factor cancels out and we obtain 
\begin{eqnarray}\label{eta}
\eta = \frac{(N + 2)\epsilon^{2}}{2(N + 8)^{2}}\left\{ 1 + \epsilon\left[ \frac{6(3N + 14)}{(N + 8)^{2}} -\frac{1}{4} \right]\right\}\equiv\eta^{(0)},
\end{eqnarray}
\begin{eqnarray}\label{nu}
&&\nu = \frac{1}{2} + \frac{(N + 2)\epsilon}{4(N + 8)} +  \frac{(N + 2)(N^{2} + 23N + 60)\epsilon^{2}}{8(N + 8)^{3}} \equiv\nu^{(0)},
\end{eqnarray}
where the exponents $\eta^{(0)}$ and $\nu^{(0)}$ are the corresponding LI ones \cite{Wilson197475}. In next section we evaluate the same critical exponents for any loop level.

\section{All-loop order critical exponents}\label{All-loop order critical exponents}

\par The results of the earlier section show that after the cancelling of the $K_{\mu\nu}$ coefficients through the momentum dependent integrals, the renormalization constants, $\beta$-function, field and composite field anomalous dimensions and the fixed point depend on $K_{\mu\nu}$ only due to $\Pi$ in a power-law form. We can see from the eqs. \ref{uhguhfgugua}-\ref{uhguhfguguja}, \ref{reewriretjgjk}-\ref{gfydsguyfsdgufa} and \ref{yagyaguhd} that the LV theory can be attained from the LI one by the substitution $g^{(0)} \rightarrow \Pi g$ in their loop quantum corrections, one power of $\Pi$ for each loop order, where $g^{(0)}$ is the LI dimensionless renormalized coupling constant. In fact, it is known that the massless LV theory can be written in terms of the LI one with the effective scaled coupling constant $g^{(0)} = \Pi g$ by applying coordinates transformations ideas \cite{0295-5075-108-2-21001} inspired in the massive case \cite{PhysRevD.84.065030}. Thus, for any loop order, we have 
\begin{eqnarray}\label{uhgufhduhufdhu}
\beta(g) = g\left( -\epsilon + \sum_{n=2}^{\infty}\beta_{n}^{(0)}\Pi^{n-1}g^{n-1}\right), 
\end{eqnarray}
\begin{eqnarray}
\gamma(g) = \sum_{n=2}^{\infty}\gamma_{n}^{(0)}\Pi^{n}g^{n},
\end{eqnarray}
\begin{eqnarray}
\gamma_{\phi^{2}}(g) = \sum_{n=1}^{\infty}\gamma_{\phi^{2}, n}^{(0)}\Pi^{n}g^{n}.
\end{eqnarray}
where $\beta_{n}^{(0)}$, $\gamma_{n}^{(0)}$ and $\gamma_{\phi^{2}, n}$ are the LI nth-loop corrections to the referred functions. The Gaussian fixed point is null as earlier in the next-to-leading level approximation. The nontrivial fixed point, evaluated from parentheses in eq. \ref{uhgufhduhufdhu}, is then $g^{*(0)} = \Pi g^{*}$, where $g^{*(0)}$ is the LI all-loop order nontrivial fixed point. The LV $\Pi$ factor cancels out in the computation of the LV critical exponents at any loop level and asserts that the LV critical exponents are equal to their LI counterparts. This final result, that the critical exponents are the same both the LV and LI theories can be completely understood only if we evoke some ideas from the research branch of phase transitions and critical phenomena. This research branch is a very traditional and rich one in providing a large amount of experimental and phenomenological evidence that completely distinct system as a fluid and ferromagnet present an identical critical behavior. In other words, they share the same set of critical exponents. This happens when they have in common their dimension $d$, $N$ and symmetry of some $N$-component order parameter if the interactions are of short-range type. As the symmetry breaking mechanism does not occur in the internal symmetry space of the field and consequently of the order parameter but in the space-time where the field is embedded, the critical exponents must not change. In terms of the renormalization group language, for asserting the influence of an operator on the critical behavior of a system, it is not suffice to including it just based on naive dimensional analysis. Its final effect on the critical exponents must be checked after the renormalization process.  

\section{Conclusions}\label{Conclusions}

In this work, we have analytically evaluated the all-loop order critical exponents for a critical thermal Lorentz-violating O($N$) self-interacting $\lambda\phi^{4}$ scalar field theory in the massless BPHZ method. For that, they have been computed in a proof by induction up to any loop level. The induction procedure was guided by an explicit calculation up to next-to-leading order. It was shown that the massles BPHZ method is simpler that its massive version in attaining the same goal, where a minimal set of diagrams was needed. We investigated the effect of the Lorentz violation in the outcome for the critical exponents and showed that the critical exponents are identical to their LI counterparts. We furnished both mathematical explanation and physical interpretation of the results as a requirement for a full understanding of them. The mathematical explanation if that the LV $K_{\mu\nu}$ coefficients can be absorbed in an effective LI theory with a scaled coupling constant by applying coordinates transformations techniques. The physical interpretation is: as the Lorentz symmetry braking mechanism does not occur in the internal symmetry space of the field but in the in which the field is embedded \cite{Aharony}, the O$(N)$ symmetry is preserved. Thus the critical exponents must not change. This work sheds light, through a formal treatment, on the study of a possible change of the critical exponents under a symmetry breaking of the order parameter.   

\appendix
\section{Integral formulas in $d$-dimensional Euclidean momentum space}\label{Integral formulas in $d$-dimensional Euclidean momentum space}

\par Considering $\hat{S}_{d} \equiv S_{d}/(2\pi)^{d} = [2^{d-1}\pi^{d/2}\Gamma(d/2)]^{-1}$ where $S_{d} = 2\pi^{d/2}/\Gamma(d/2)$ is the unit $d$\hyp{}dimensional sphere area, we have

\begin{eqnarray}
&&\int \frac{d^{d}q}{(2\pi)^{d}} \frac{1}{(q^{2} + 2pq + M^{2})^{\alpha}} = \hat{S}_{d}\frac{1}{2}\frac{\Gamma(d/2)}{\Gamma(\alpha)}\frac{\Gamma(\alpha - d/2)}{(M^{2} - p^{2})^{\alpha - d/2}},
\end{eqnarray}

\begin{eqnarray}
&&\int \frac{d^{d}q}{(2\pi)^{d}} \frac{q^{\mu}}{(q^{2} + 2pq + M^{2})^{\alpha}} = -\hat{S}_{d}\frac{1}{2}\frac{\Gamma(d/2)}{\Gamma(\alpha)}\frac{p^{\mu}\Gamma(\alpha - d/2)}{(M^{2} - p^{2})^{\alpha - d/2}},
\end{eqnarray}

\begin{eqnarray}
&&\int \frac{d^{d}q}{(2\pi)^{d}} \frac{q^{\mu}q^{\nu}}{(q^{2} + 2pq + M^{2})^{\alpha}} = \hat{S}_{d}\frac{1}{2}\frac{\Gamma(d/2)}{\Gamma(\alpha)} \nonumber \\&&   \times \left[ \frac{1}{2}\delta^{\mu\nu}\frac{\Gamma(\alpha - 1 - d/2)}{(M^{2} - p^{2})^{\alpha - 1 - d/2}} + p^{\mu}p^{\nu}\frac{\Gamma(\alpha - d/2)}{(M^{2} - p^{2})^{\alpha - d/2}} \right],
\end{eqnarray}

\begin{eqnarray}
&&\int \frac{d^{d}q}{(2\pi)^{d}} \frac{q^{\mu}q^{\nu}q^{\rho}}{(q^{2} + 2pq + M^{2})^{\alpha}} = - \hat{S}_{d}\frac{1}{2}\frac{\Gamma(d/2)}{\Gamma(\alpha)}  \nonumber \\&&   \times\left[\frac{1}{2}[\delta^{\mu\nu}p^{\rho} + \delta^{\mu\rho}p^{\nu} + \delta^{\nu\rho}p^{\mu}]\frac{\Gamma(\alpha - 1 - d/2)}{(M^{2} - p^{2})^{\alpha - 1 - d/2}} + p^{\mu}p^{\nu}p^{\rho}\frac{\Gamma(\alpha - d/2)}{(M^{2} - p^{2})^{\alpha - d/2}} \right],
\end{eqnarray}

\begin{eqnarray}
&&\int \frac{d^{d}q}{(2\pi)^{d}} \frac{q^{\mu}q^{\nu}q^{\rho}q^{\sigma}}{(q^{2} + 2pq + M^{2})^{\alpha}} = \hat{S}_{d}\frac{1}{2}\frac{\Gamma(d/2)}{\Gamma(\alpha)}  \nonumber \\&&  \times\left[\frac{1}{4}[\delta^{\mu\nu}\delta^{\rho\sigma} + \delta^{\mu\rho}\delta^{\nu\sigma} +\delta^{\mu\sigma}\delta^{\nu\rho}]\frac{\Gamma(\alpha - 2 - d/2)}{(M^{2} - p^{2})^{\alpha - 2 - d/2}}  \right.  \nonumber \\  &&\left. + \frac{1}{2}[\delta^{\mu\nu}p^{\rho}p^{\sigma} + \delta^{\mu\rho}p^{\nu}p^{\sigma} + \delta^{\mu\sigma}p^{\nu}p^{\rho} + \delta^{\nu\rho}p^{\mu}p^{\sigma} +\delta^{\nu\sigma}p^{\mu}p^{\rho} +\delta^{\rho\sigma}p^{\mu}p^{\nu}]\frac{\Gamma(\alpha - 1 - d/2)}{(M^{2} - p^{2})^{\alpha - 1 - d/2}}   \right.  \nonumber \\  &&\left. + p^{\mu}p^{\nu}p^{\rho}p^{\sigma}\frac{\Gamma(\alpha - d/2)}{(M^{2} - p^{2})^{\alpha - d/2}} \right].
\end{eqnarray}

\bibliography{apstemplate}

\end{document}